\documentclass{iopart}

\usepackage{iopams}
\usepackage{epsf}
\usepackage{graphicx}
\usepackage{color}
\usepackage{pstcol}
\usepackage[all]{xy}

\definecolor{light-red}{rgb}{1,0.8,0.8}
\definecolor{light-blue}{rgb}{0.5,0.9,1}
\definecolor{dark-yellow}{rgb}{1,0.8,0}
\newpsobject{showgrid}{psgrid}{subgriddiv=1,griddots=10,gridlabels=6pt}

\begin{document}

\title[OPTIS - satellite--based test of Special and General Relativity]{OPTIS -- a satellite--based test of Special and General Relativity}

\author{Claus L{\"a}mmerzahl\dag\,, Hansj\"org Dittus\ddag\,, Achim Peters\S\,, and Stephan Schiller\dag}

\address{\dag\ Institute for Experimental Physics, Heinrich--Heine--University D\"usseldorf, 40225 D\"usseldorf, Germany }

\address{\ddag\ ZARM, University of Bremen, 28359 Bremen, Germany}

\address{\S\ Department of Physics, University of Konstanz, 78457 Konstanz, Germany}

\begin{abstract}
A new satellite based test of Special and General Relativity is proposed. For the Michelson--Morley experiment we expect an improvement of at least three orders of magnitude, and for the Kennedy--Thorndike experiment an improvement of more than one order of magnitude. Furthermore, an improvement by two orders of the test of the universality of the gravitational red shift by comparison of an atomic clock with an optical clock is projected. -- The tests are based on ultrastable optical cavities, an atomic clock and a comb generator. 
\end{abstract}




\section{Introduction and Motivation}

Special Relativity (SR) and General Relativity (GR) are at the basis of our understanding of space and time and thus are fundamental for the formulation of physical theories. 
Without SR we cannot explain the phenomena in high energy physics and in particle astrophysics, without GR there is no understanding of the gravitational phenomena in our solar system, of the dynamics of galaxies and of our universe, and, finally, of the physics of black holes. Both theories are linked by the fact that the validity of ST is necessary for GR. Due to the overall importance of these theories, a persistent effort to improve the  experimental tests of their foundations is mandatory. 
Modern tests of SR and GR using ultrastable oscillators have been performed on earth \cite{BrilletHall79,HilsHall90,Turneaureetal83}, and an experiment is planned on the International Space Station \cite{Buchmanetal98}. 
 
Furthermore, new results from quantum gravity theories predict small deviations from SR and GR giving additional motivation to improve tests on SR and GR. For example, loop gravity and string theory predict modifications of the Maxwell equations \cite{GambiniPullin99,EllisMavromatosNanopoulos99}. These modifications lead to an anisotropic speed of light and to a dispersion in vacuum, thus violating the postulates of SR. Quantum gravity also predicts, besides a violation of the Weak Equivalence Principle \cite{DamourPolyakov94}, a violation of the universality of the gravitational red shift \cite{Damour97,Damour00}. 
Although the amount of SR--violations predicted by quantum gravity are too small to be in the range of experimental capabilities in the near future, these predictions open up the window for violations for such basic principles. 
Since we do not know the ''true'' quantum gravity theory we also do not know the ''true'' parameters of the theory and therefore the predicted range of violations of SR and GR is no final prediction, but merely a hint. 

\section{Overview of OPTIS}

\begin{figure}[t!]
\psset{unit=1.12cm}
\hspace*{-4mm}\begin{pspicture}(-5,-3)(7,5)
\psframe[fillstyle=solid,fillcolor=black](-4.6,-3.5)(7,5)
\pscircle[fillstyle=solid,fillcolor=white](-3,0){0.47}
\rput(-3,0){\includegraphics[width=1cm]{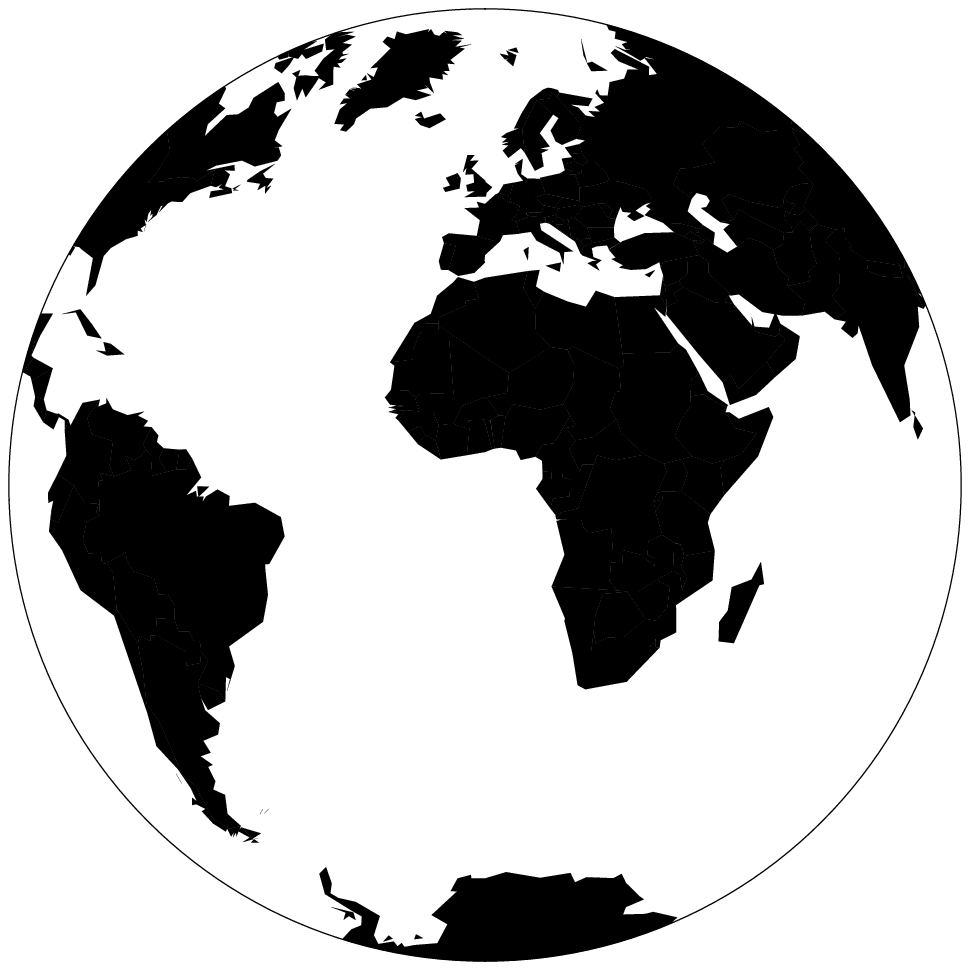}}
\rput{30}(-0.5,1.5){\psellipse[linecolor=white](0,0)(4.5,1.7)}
\rput(-3,-1.5){\color{white}\footnotesize perigee $\sim$ 10000 km}
\rput(3.7,4.4){\color{white}\footnotesize apogee $\sim$ 40000 km}
\psline[linecolor=white](-0.5,3.4)(4.8,2.2)
\psline[linecolor=white](-0.5,3.3)(1.5,-1.7)
\rput{90}(-1.7,3.3){\psellipse[linecolor=white](0,0.3)(0.6,0.2)
\psframe[linestyle=none,fillstyle=solid,fillcolor=black](-0.3,0.3)(0.3,1)
\psline[linecolor=white]{->}(0.4,0.42)(0.3,0.45)}
\rput{90}(-0.5,3.3){\psset{unit=0.2cm}
\psline[fillstyle=solid,fillcolor=light-blue](-1.5,0.5)(-0.5,1.5)(-3.5,1.5)(-4.5,0.5)(-1.5,0.5)
\psellipse[fillstyle=solid,fillcolor=red](0,0)(1,0.5)
\psframe[linestyle=none,fillstyle=solid,fillcolor=red](-1,0)(1,2)
\psline(-1,0)(-1,2)
\psline(1,0)(1,2)
\psellipse[fillstyle=solid,fillcolor=red](0,2)(1,0.5)
\psline[fillstyle=solid,fillcolor=light-blue](0.5,0.5)(1.5,1.5)(4.5,1.5)(3.5,0.5)(0.5,0.5)
\psline[linecolor=white]{->}(0,2)(0,9)
}
\rput(-1.4,3.5){\color{white}$\omega$}
\psline[linecolor=white]{->}(-2,4.3)(-4,4.3)
\rput(-3,4.6){\color{white}to sun}
\pscircle[fillstyle=solid,fillcolor=light-blue,linewidth=2pt,linecolor=red](4,-0.45){2.8}
\rput(4.75,-0.15){
\psset{unit=0.56cm}
\psline[linewidth=2pt,linecolor=blue,fillstyle=solid,fillcolor=white](3.2,4.2)(3.2,-1.2)(1.2,-3.2)(-4.2,-3.2)(-6.2,-1.2)(-6.2,4.2)(-4.2,6.2)(1.2,6.2)(3.2,4.2)
\psframe[linestyle=none,fillstyle=solid,fillcolor=blue](0,-0.8)(0.8,0)
\psframe[linestyle=none,fillstyle=solid,fillcolor=blue](1.2,-0.8)(2,0)
\psframe[linestyle=none,fillstyle=solid,fillcolor=blue](0,-2)(0.8,-1.2)
\psframe[linestyle=none,fillstyle=solid,fillcolor=blue](1.2,-2)(2,-1.2)
\psline[linecolor=blue](0,-0.8)(0,0)(0.8,0)  
\psline[linecolor=blue](1.2,0)(2,0)(2,-0.8)
\psline[linecolor=blue](0,-1.2)(0,-2)(0.8,-2)  
\psline[linecolor=blue](1.2,-2)(2,-2)(2,-1.2)
\psline[linewidth=2pt](2,-1.2)(2,-0.8)
\psline[linewidth=2pt](0.8,-2)(1.2,-2)
\psline[linewidth=2pt](0,-1.2)(0,-0.8)
\psline[linewidth=2pt](0.8,0)(1.2,0)
\psframe[fillstyle=solid,fillcolor=red](-5.5,-1.3)(-4,-0.7)
\psline(-4,-0.9)(-4,-0.7)(-5.5,-0.7)(-5.5,-1.3)(-4,-1.3)(-4,-1.1)
\psframe[fillstyle=solid,fillcolor=red](0.7,3)(1.3,4.5)
\psline(0.9,3)(0.7,3)(0.7,4.5)(1.3,4.5)(1.3,3)(1.1,3)
\psframe[fillstyle=solid,fillcolor=yellow](-6,0.7)(-4.5,1.3)
\psframe[fillstyle=solid,fillcolor=light-red](-6,2.7)(-4.5,3.3)
\psline[doubleline=true]{->}(-5.25,1.3)(-5.25,2.7)
\put(-1,4.2){
\pswedge[fillstyle=solid,fillcolor=dark-yellow,linestyle=none](0,0){0.3}{0}{180}
\psarc(0,0){0.3}{0}{180}
\psline[doubleline=true]{->}(0,0.3)(0,1.2)
}
%
\rput{90}(-1.9,1){
\pswedge[fillstyle=solid,fillcolor=dark-yellow,linestyle=none](0,0){0.3}{0}{180}
\psarc(0,0){0.3}{0}{180}}
\psline[linewidth=3pt,linecolor=gray](0.5,0.5)(1.5,1.5)
\psline[linewidth=3pt,linecolor=gray](-1.5,-1.5)(-0.5,-0.5)
\psline[linewidth=3pt,linecolor=gray](-0.5,0.5)(-1.5,1.5)
\psline[linewidth=3pt,linecolor=gray](-1.5,2.5)(-0.5,3.5)
\psline[linewidth=1pt,linecolor=red]{->}(-4,-1)(-2.5,-1)
\psline[linewidth=1pt,linecolor=red]{->}(-2.5,-1)(1.6,-1)
\psline[linewidth=1pt,linecolor=red]{->}(2,-1)(0.4,-1)
\psline[linewidth=1pt,linecolor=red]{->}(-1,-1)(-1,0)
\psline[linewidth=1pt,linecolor=red]{->}(-1,0)(-1,2)
\psline[linewidth=1pt,linecolor=red]{->}(-1,2)(-1,4.2)
\psline[linewidth=1pt,linecolor=red]{->}(-4.5,3)(-2.75,3)
\psline[linewidth=1pt,linecolor=red](-2.75,3)(-1,3)
\psline[linewidth=1pt,linecolor=red]{->}(1,3)(1,1.25)
\psline[linewidth=1pt,linecolor=red]{->}(1,2.5)(1,-1.6)
\psline[linewidth=1pt,linecolor=red]{->}(1,-2)(1,-0.4)
\psline[linewidth=1pt,linecolor=red]{->}(1,1)(0,1)
\psline[linewidth=1pt,linecolor=red]{-}(0,1)(-1,1)
\psline[linewidth=1pt,linecolor=red]{->}(-1,1)(-1.9,1)
\psline[doubleline=true]{->}(-2.2,1)(-3,1)
\psline[doubleline=true]{->}(-4.5,1)(-3.64,1)
\psline[doubleline=true]{-}(-3.34,1.3)(-3.34,2.9)
\psline[doubleline=true]{->}(-3.34,3.1)(-3.34,5.4)
\rput(-3.34,1){\LARGE $\otimes$}
%
\rput(-4.5,-1.6){\footnotesize laser 1}
\rput{90}(1.6,3.75){\footnotesize laser 2}
\rput(0,-2.3){\txt\footnotesize{crossed}}
\rput(0,-2.7){\txt\footnotesize{resonators}}
\rput(-2.2,5.7){\footnotesize to data recording}
\rput{90}(2.7,1.5){\footnotesize thermal isolation}
\rput(-5.1,0.4){\footnotesize atomic }
\rput(-5.2,-0.1){\footnotesize clock}
\rput(-5,4.1){\footnotesize comb}
\rput(-4.9,3.6){\footnotesize generator}
}
\end{pspicture}
\vspace*{1mm}
\caption{Scheme of OPTIS: The science payload of the satellite mainly consists of two crossed resonators to which two lasers are locked, an atomic clock and an optical comb generator. 
The orbit of the satellite is highly elliptic. \label{OPTISPrinzip}}
\end{figure}

By means of the proposed mission OPTIS an improvement of three tests of SR and GR by up to three orders of magnitude is projected:

\vspace*{-2mm}
\begin{center}
\noindent\begin{tabular}{llc}
Experiment & present accuracy & projected accuracy \\  
[1ex]\hline
\rule[-5pt]{0pt}{20pt}Michelson--Morley--experiment (MM) & $\delta_\vartheta c/c \leq  
3 \times 10^{-15}$ \hfill \cite{BrilletHall79} & $\leq 10^{-18}$ \\
Kennedy--Thorndike--experiment (KT) & $\delta_v c/c \leq 2 \times 10^{-13}$  
\hfill \cite{HilsHall90} & $\leq 10^{-15}$ \\
universality of gravitat.\ red shift (LPI) & $\Delta\alpha \leq 2 \times 10^{-2}$ \hfill \cite{Godoneetal95} & $\Delta\alpha \leq 10^{-4}$ \\ [1ex] \hline
\end{tabular}
\end{center}

\smallskip
The main features of the OPTIS mission can be seen in Fig.\ref{OPTISPrinzip}: A spinning drag--free satellite orbits the earth. The satellite payload consists of two lasers, two orthogonal optical cavities, a femtosecond laser comb generator, and an atomic clock. The two cavities are used for the MM experiment which searches for differences in the velocity of light in different directions. The atomic clock represents an independent clock of different physical nature. A comparison between the atomic clock and an optical cavity can be performed by means of the comb generator. A search for a dependence of their frequency ratio with respect to a change of the velocity of the satellite or with respect to a change of the gravitational potential amounts to a KT test and to a test of the universality of the gravitational red shift, respectively. 

\section{Theoretical description and present status}

\begin{figure}[t!]
\begin{center}
\psset{unit=0.7cm}
\begin{pspicture}(1,-2)(9,7)
\multips(5,1.4)(0,0.02){10}{\psellipse[linecolor=black,fillstyle=solid,fillcolor=lightgray](0,0)(4,1)}
\psline[linecolor=blue,fillstyle=solid,fillcolor=light-blue](4.4,1.6)(5.2,1.4)(5.7,2.3)(4.9,2.5)(4.4,1.6)
\psline[linecolor=blue,fillstyle=solid,fillcolor=blue](5.2,1.4)(5.7,2.3)(5.7,1.7)
\psline[linewidth=1pt,linecolor=red]{->}(2.7,1.3)(5,2)
\psline[linewidth=1pt,linecolor=red]{->}(5,2)(5,4)
\psline[linewidth=1pt,linecolor=red]{-}(5,4)(5,5)
\multips(3.7,1.6)(-0.05,-0.015){20}{\psellipse[linecolor=red,fillstyle=solid,fillcolor=black](0,0)(0.25,0.4)}
\psline[linewidth=0.3cm](5,-1)(5,0.36)
\put(5,5){\pswedge[fillstyle=solid,fillcolor=yellow,linestyle=none](0,0){0.3}{0}{180}
\psarc(0,0){0.3}{0}{180}
\psbezier(0,0.3)(0,1.8)(0.7,-0.2)(0.7,1.3)}
\rput(7.3,5.1){\txt{measurement of \\ frequency $\nu$}}
\rput(6.5,1.8){mirror}
\rput(3,2.8){cavity}
\rput(-0.3,1.2){turn table}
\rput(5,0.8){\pscurve[linewidth=2pt]{->}(0.5,-0.992157)(1,-0.968246)(1.5,-0.927025)(2,-0.866025)(2.5,-0.780625)(3,-0.661438)(3.5,-0.484123)(3.7,-0.379967)(3.9,-0.222205)}
\rput(8.3,-0.2){$\vartheta$}
\psline[linewidth=2pt]{->}(4,-2)(7,-1)
\rput(7,-1.4){$\mbox{\boldmath$v$}$}
\end{pspicture}
\end{center}
\vspace*{-1mm}
\caption{Basic setup for a cavity experiment testing the spatial isotropy and the independence of the speed of light from the velocity $v$ of the laboratory. 
If SR is valid, then the measured frequency is independent of $\vartheta$ and $v$. \label{Fig:SchemeTest}}
\end{figure}
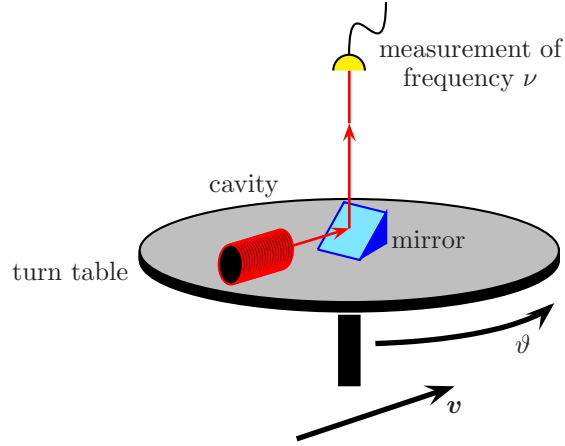

The most precise experiments testing the constancy of the speed of light use cavities. The wave vector magnitude $k$ of an electromagnetic wave in a cavity of length $L$ is given by $k = n \pi/L$ and the frequency $\nu$ of an outcoupled wave by $\nu = c\, k$. 
If the velocity of light depends on the orientation of the cavity and on the velocity $v$ of the laboratory, $c = c(\vartheta, v)$, so will the frequency $\nu = \nu(\vartheta, v) = c(\vartheta, v)\, k$. 
Fig.\ref{Fig:SchemeTest} shows a schematic setup for a search of an orientation and velocity dependence of the frequency. Since SR is based on an orientation and velocity {\it in}dependent speed of light, the search for an orientiation and velocity dependence of the frequency amounts to a test of SR.

According to common test theories \cite{Robertson49,MansouriSexl77}, the orientation and velocity dependence of the velocity of light is parametrized according to 
\begin{equation}
c(\vartheta, v) = c\left(1  + A \; \frac{v^2}{c^2} \sin^2\vartheta + B \; \frac{v^2}{c^2} + {\cal O}(v^4/c^4)\right)\, . \label{GenAnsatz}
\end{equation}
This means that all anomalous terms vanish for vanishing velocity $v$. Here $\vartheta$ is the angle between the velocity with respect to the cosmic preferred frame $\mbox{\boldmath$v$}$ and the cavity axis. 
In addition, an expansion with respect to $v^2/c^2$ has been used, for simplicity. 
If Special Relativity is valid, then $A = B = 0$. 
The parameters $A$ and $B$ in the Mansouri--Sexl and Robertson test theories are given by

\medskip
\renewcommand{\arraystretch}{1.2}
\begin{tabular}{|l|ll|} \hline
parameter & Robertson test theory & Mansouri--Sexl test theory \\ \hline
$A$ & $|1 - g_1(v)/g_2(v)|$ & $\beta + \delta - \frac{1}{2}$ \\ 
$B$ & $|1 - g_1(v)/g_0(v)|$ & $\alpha - \beta + 1$ \\ \hline
\end{tabular}
\renewcommand{\arraystretch}{1}

\subsection{Isotropy of space}

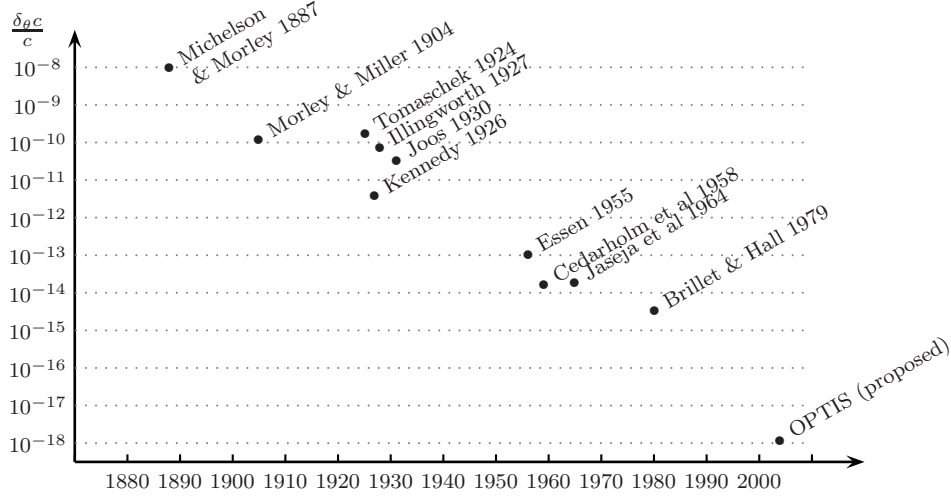
\begin{figure}[t]
\begin{center}
\psset{xunit=0.7cm,yunit=0.5cm}
\begin{pspicture}(0,-1)(16,11)
\psline[linewidth=1pt]{->}(1,-1.5)(16,-1.5)
\psline[linewidth=1pt]{->}(1,-1.5)(1,10)
\rput(0.1,10){$\frac{\delta_\theta c}{c}$}
\multips(1,-1.5)(1,0){15}{\psline(0,0)(0,0.2)}
\rput(2,-2){\footnotesize 1880}
\rput(3,-2){\footnotesize 1890}
\rput(4,-2){\footnotesize 1900}
\rput(5,-2){\footnotesize 1910}
\rput(6,-2){\footnotesize 1920}
\rput(7,-2){\footnotesize 1930}
\rput(8,-2){\footnotesize 1940}
\rput(9,-2){\footnotesize 1950}
\rput(10,-2){\footnotesize 1960}
\rput(11,-2){\footnotesize 1970}
\rput(12,-2){\footnotesize 1980}
\rput(13,-2){\footnotesize 1990}
\rput(14,-2){\footnotesize 2000}
\rput(0.3,-1){\footnotesize $10^{-18}$}
\psline[linestyle=dotted,linecolor=gray](1,-1)(15,-1)
\rput(0.3,0){\footnotesize $10^{-17}$}
\psline[linestyle=dotted,linecolor=gray](1,0)(15,0)
\rput(0.3,1){\footnotesize $10^{-16}$}
\psline[linestyle=dotted,linecolor=gray](1,1)(15,1)
\rput(0.3,2){\footnotesize $10^{-15}$}
\psline[linestyle=dotted,linecolor=gray](1,2)(15,2)
\rput(0.3,3){\footnotesize $10^{-14}$}
\psline[linestyle=dotted,linecolor=gray](1,3)(15,3)
\rput(0.3,4){\footnotesize $10^{-13}$}
\psline[linestyle=dotted,linecolor=gray](1,4)(15,4)
\rput(0.3,5){\footnotesize $10^{-12}$}
\psline[linestyle=dotted,linecolor=gray](1,5)(15,5)
\rput(0.3,6){\footnotesize $10^{-11}$}
\psline[linestyle=dotted,linecolor=gray](1,6)(15,6)
\rput(0.3,7){\footnotesize $10^{-10}$}
\psline[linestyle=dotted,linecolor=gray](1,7)(15,7)
\rput(0.3,8){\footnotesize $10^{-9}$}
\psline[linestyle=dotted,linecolor=gray](1,8)(15,8)
\rput(0.3,9){\footnotesize $10^{-8}$}
\psline[linestyle=dotted,linecolor=gray](1,9)(15,9)
\rput{30}(2,8){\put(0,0){\txt\footnotesize{$\bullet$ Michelson \\ \textcolor{white}{...........}\& Morley 1887}}}
\rput{30}(4.4,7){\put(0,0){\footnotesize $\bullet$ Morley \& Miller 1904}}
\rput{30}(6.4,7.2){\put(0,0){\footnotesize $\bullet$ Tomaschek 1924}}
\rput{30}(6.6,5.5){\put(0,0){\footnotesize \footnotesize $\bullet$ Kennedy 1926}}
\rput{30}(6.7,6.8){\put(0,0){\footnotesize $\bullet$ Illingworth 1927}}
\rput{30}(7,6.5){\put(0,0){\footnotesize $\bullet$ Joos 1930}}
\rput{30}(9.5,4){\put(0,0){\footnotesize $\bullet$ Essen 1955}}
\rput{30}(9.8,3.2){\put(0,0){\footnotesize $\bullet$ Cedarholm et al 1958}}
\rput{30}(10.4,3.2){\put(0,0){\footnotesize $\bullet$ Jaseja et al 1964}}
\rput{30}(11.9,2.5){\put(0,0){\footnotesize $\bullet$ Brillet \& Hall 1979}}
\rput{30}(14.3,-1){\put(0,0){\footnotesize $\bullet$ OPTIS (proposed)}}
\end{pspicture}
\end{center}
\medskip
\caption{Improvements of optical tests of the isotropy of space. \label{Fig:HistoryMM}}
\end{figure}

In order to test the isotropy of space, or the isotropy of the velocity of light, one has to mount the cavity on a turn table and look for a variation of the frequency as the turn table is rotated. 
This setup has been used in experiments since 1955, see Fig.\ref{Fig:HistoryMM}. In terms of the relative variation of the velocity of light, of the Robertson parameter, and of the Mansouri--Sexl parameter, the most accurate result is \cite{BrilletHall79}
\begin{equation}
\frac{\delta_\vartheta c}{c} \leq 3 \times 10^{-15}\; , \quad \left|1 - \frac{g_2(v)}{g_1(v)}\right| \leq 5 \times 10^{-15}\; , \quad \left|\beta + \delta - \frac{1}{2}\right|  \leq 2.5 \times 10^{-15} \, . 
\end{equation}
For more interpretation see \cite{LH01}. 

\subsection{Independence of the velocity of light from the velocity of the laboratory}

A hypothetical dependence of the velocity of light on the velocity $v$ of the laboratory can be tested by changing the velocity of the cavity and looking for a variation of the frequency. 
In earth based experiments, the rotation of the earth around its axis ($v = v_0 \pm 300\;\hbox{m/s}$) or around the sun ($v = v_0 \pm 30\;\hbox{km/s}$) can be used, where $v_0 = 377\;\hbox{km/s}$ is the velocity with respect to the cosmic microwave background, the cosmologically preferred frame.   
Because of technical reasons only the rotation of the earth around its own axis has been used so far. In terms of the same parameters as above, the most accurate result is  \cite{HilsHall90}
\begin{equation}
\frac{\delta_v c}{c} \leq 2 \times 10^{-13}\; , \quad \left|1 - \frac{g_1(v)}{g_0(v)}\right| \leq 2 \times 10^{-13}\; , \quad \left|\alpha - \beta + 1\right| \leq 7 \times 10^{-7} \, .
\end{equation}

\subsection{Gravitational red shift}

In the framework of Einstein's GR, the comparison of two identical clocks of frequency $\nu_0$ located in different gravitational potentials $U(x_1)$ and $U(x_2)$ yields $\nu(x_2) = \left(1 + (U(x_2) - U(x_1))/c^2\right) \nu(x_1)$. 
The gravitational red shift does not depend on the type of clock. 
This is the universality of the gravitational red shift, an aspect of Local Position Invariance. 

If Einstein's theory is not correct, then the red shift may depend on the clock $\nu(x_2) = \left(1 + \alpha_{\rm clock} (U(x_2) - U(x_1))/c^2\right) \nu_(x_1)$ with $\alpha_{\rm clock} \neq 1$. 
In Einstein's GR $\alpha_{\rm clock} = 1$. 
If two different clocks are displaced together in a gravitational potential, a relative frequency shift $\Delta\nu_{\rm clock 1}/\nu_{01} - \Delta\nu_{\rm clock 2}/\nu_{02} = \left(\alpha_{\rm clock 1} - \alpha_{\rm clock 2}\right) \Delta U/c^2$ may occur, which is proportional to the difference of the gravitational potential relative to the uinitial position. 

For the hydrogen maser $|\alpha_{\rm H-maser} - 1| \leq 10^{-4}$ (GP-A experiment, \cite{VessotLevine80}), verifying the gravitational red shift. 
For a test of the universality of the gravitational red shift by means of a comparison between a microwave cavity and an atomic Cesium clock, the best result is $|\alpha_{\rm atom} - \alpha_{\rm cavity}| \leq 10^{-2}$ \cite{Turneaureetal83}.

\subsection{Advantages of a satellite--based test}

The advantages of a sattelite mission for doing experiments for testing SR and GR are the following:

\begin{itemize}

\item[{(i)}] The high orbital velocity (see Eq.(\ref{GenAnsatz}). For the proposed (elliptic) orbit, $v$ varies between $+ 7 \;\hbox{km/s}$ and $- 4 \; \hbox{km/s}$ over half the orbit period $T_{\rm orbit}/2 \sim 7 \;\hbox{h}$. 
This value is 20 times larger than the change of the velocity of the earth's surface over a period of 12 h.

\item[{(ii)}] The shorter period $T_{\rm orbit}$ of the velocity modulation as compared to 24 h on earth. It relaxes the requirements on the optical resonators and is directly relevant for the KT and LPI tests. 
It is indirectly useful also for the elimination of systematic effects in the MM test.

\item[{(iii)}] The difference of the gravitational potential in the highly eccentric orbit is $\Delta U/c^2 \sim 3 \times 10^{-10}$. This is about three orders of magnitude larger than the difference of the potential of the sun which an earth--bound observer experiences. This is relevant for the LPI test. 

\item[{(iv)}] The microgravity environment minimizes distortions of the optical resonators. This is relevant for the MM test. 

\item[{(v)}] A variable spin frequency of the satellite permits elimination of sytematic effects.

\item[{(vi)}] Long integration times (longer than 6 months). 

\end{itemize}

\section{The science payload}

In this and the following section we discuss the science payload of 
the satellite, the relevant performance specifications and the 
resulting requirements for the orbit and the satellite bus structure.

In order to perform the MM test, the satellite has to spin around its axis. A typical rotation period is $\tau_{\rm MM} = 2\pi/\omega \sim 100 - 1000\,\hbox{s}$. 
For the elimination of systematic errors, the rotation frequency $\omega$ can be varied. 
For the KT test the timescale is the orbit time $\tau_{\rm KT} = T_{\rm orbit} \sim 10^5 \,\hbox{s}$. 

The main subsystems of the experimental payload are optical 
resonators, ultra stable lasers, an optical frequency comb generator 
and an ultra stable microwave oscillator. These components are 
interconnected and supplemented by the locking and 
stabilization electronics, the optical bench, the drag free control 
system (discussed in the next section), and an advanced thermal control 
system. 

\subsection{Thermal control system}

The LISA pre--phase A report \cite{LISAPreA} showed that a 3--stage passive 
thermal isolation is sufficient to achieve a level of thermal 
fluctuations below $10^{-7} \,{\rm K} / \sqrt{\rm Hz}$ at $\tau 
= 1000 \;{\rm s}$. This performance (assuming no shadow phases) would 
be sufficient for the OPTIS random noise requirements.
To supress the fluctuations 
correlated with the rotation of the satellite, an improvement of the thermal stability by one order of 
magnitude would be required. This could be achieved either by an 
additional stage of passive isolation, or by adding an active temperature 
stabilization.  

\subsection{Optical resonators}

The optical resonators are the central part of the experimental setup. 
In the baseline configuration, two crossed standing wave resonators 
are implemented by optically contacting four highly reflective mirrors to 
a monolithic spacer block with two orthogonal holes made from a low--expansion glass ceramic (ULE, ZERODUR). With a technically feasible 
finesse of $2.5\times 10^5$ and a length of $10\,{\rm cm}$ each, 
these resonators should exhibit linewidths of $6\,{\rm kHz}$.

Low expansion glass ceramics are designed for a minimal 
thermal expansion coefficient ($\lesssim 10^{-9} / {\rm K}$) at room 
temperature and can be manufactured in many different shapes and 
dimensions. Resonators made from these materials are well suited for 
laser stabilization, although aging effects cause a continuous 
shrinking of the material and thereby frequency drifts of typically $5 
- 50 \,{\rm kHz}/{\rm day}$.

The monolithic construction of the resonator block serves to strongly 
reduce the effect of shrinking and of 
temperature fluctuations on the MM experiment: A high degree of 
common mode rejection (two orders of magnitude) in the differential frequency measurement of the lasers locked to the resonators is expected.

For the KT and the redshift experiment, on the other hand, the aging related frequency drift is critical. It has to be modelled well enough to keep the unpredictable residual part below $2 \times 10^{-13}$ (i.e. $50\,{\rm Hz}$) over the signal half period $T_{\rm orbit}/2 = 7\;\hbox{h}$. This is feasible \cite{HilsHall90,Youngetal99}.

For the projected measurement sensitivity the temperature stability requirements are as follows: a level $\Delta T(\tau_{\rm MM}) \leq 20\;\mu\hbox{K}$ for random fluctuations over the spin period time scale and a level $\Delta T_{\rm sys} \leq 100 \,{\rm nK}$ for 
a temperature modulation correlated with the rotation of the 
satellite. For the KT as well as for the redshift experiment the requirements are $\Delta T(\tau_{\rm KT}) \leq 200\;\mu\hbox{K}$ for random temperature fluctuations on a time scale of the orbit period and $\Delta T_{\rm sys} \leq 10\;\mu\hbox{K}$ for temperature modulation correlated with the orbital motion. 

The length of the reference resonators is affected by accelerations, 
with a typical sensitivity of $1\,{\rm nm} / {\rm g}$ for a resonator 
length of $25 \,{\rm cm}$ \cite{Youngetal99}. This leads to a 
requirement of $6 \times 10^{-8} \,{\rm g}$ for random residual 
accelerations and $3 \times 10^{-10} \,{\rm g}$ for residual 
accelerations correlated with the rotation of the satellite. The 
drag-free control system (see next section) is designed to meet these 
requirements.

\subsection{Ultra stable lasers}

The lasers used for the OPTIS mission should have high intrinsic 
frequency stability, narrow linewidth and high intensity stability. 
These requirements are best fulfilled by diode--pumped monolithic Nd:YAG lasers, 
which are also used in gravity wave detectors (GEO600, LIGO, VIRGO, 
LISA) and many other high precision experiments \cite{Seeletal97}. Such lasers are already available in space --qualified versions.

\subsection{Electronics for frequency and intensity stabilization}

Locking the lasers to the reference resonators using the 
Pound--Drever--Hall frequency modulation method requires fast, low noise 
photodetectors and an optimized electronic control system. The 
intrinsic noise of the photodetectors has to be sufficiently low to 
allow shot noise limited detection. The residual amplitude 
fluctuations of the lasers have to be actively suppressed. To prevent 
thermally induced length changes of the resonators by absorbed laser 
radiation, the intensity of the laser beams has to be actively 
stabilized to a relative level of $10^{-4} / \sqrt{\rm Hz}$ in the frequency 
range $0.1 \,{\rm mHz}$ to  $10 \,{\rm Hz}$.

\subsection{Optical bench}

The whole optical setup should be stable and isolated from vibrations 
to prevent frequency fluctuations caused by vibration induced Doppler 
shifts. Even very small displacements (less 
then $1\,{\rm \mu m}$) of the laser beams relative to the reference 
resonators are known to cause substantial frequency shifts \cite{Storzetal98}. These 
requirements can be fulfilled by using a well designed monolithic 
optical bench on which all optical components are stably mounted as 
close together as possible.

\subsection{Atomic clock}

The KT and the universality of red--shift experiments require an independent frequency reference. This reference should 
be based on the difference of two energy levels in an atomic or 
molecular system. Atomic clocks based on hyperfine transitions in 
cesium or rubidium atoms are suited for this task. They are 
available in space qualified versions with relative instabilities of 
better than $1 \times 10^{-13}$ for the relevant time scale $\tau_{\rm KT}$ of several 
hours.

The atomic clock can also serve as the reference for the microwave 
synthesizer required to mix down the beat signal between the two stabilized lasers in the MM experiment from typically 1 GHz to a lower frequency for data acquisition and analysis. The required relative instability 
$\lesssim 10^{-12}$ on the timescale $\tau_{\rm MM}$ of $10 - 1000\,{\rm s}$ is thereby satisfied. 

\subsection{Optical comb generator}

For the frequency comparison between the atomic clock and the lasers 
stabilized to the reference resonators it is necessary to multiply the 
microwave output of the atomic clock ($\sim 10^{10} \,{\rm Hz}$) into 
the optical range ($\sim 10^{15} \,{\rm Hz}$). Thanks to recent important 
progress in the field of frequency metrology, this can now be done 
reliably amd simply by using femtosecond optical comb generators \cite{Diddams,Reichert}. 
Here the repetition rate of a mode--locked femtosecond laser of  
$\sim 1 \,{\rm GHz}$ is locked to an atomic clock. Its optical spectrum (a comb of frequencies spaced at exactly the repetition rate) is broadened to more than one octave by passing the pulses through a special optical fiber. By measuring and stabilizing the 
beatnote between the high frequency part of the comb and the frequency 
doubled low frequency part, it is then possible to determine the 
absolute frequency of each component of the comb relative to the 
atomic clock. In a final step the frequency of the cavity  
stabilized lasers is then compared to the closest frequency component 
of the comb by measuring their beatnote with a fast photodetector.

Compact diode pumped comb generators with very low power consumption, as required for space applications, are already under development. 

\section{Orbit and satellite}

\subsection{The orbit and orbit requirements}

The experimental requirements define an optimum mission profile. In particular, the requirements of 
attitude control, maximum residual acceleration, and temperature 
stability result in the following specifications:

\begin{itemize}

\item The satellite needs a drag-free attitude and orbit control system 
for all 6 degrees of freedom. Thrust control must be possible down to 
$0.1 \; \mu$N by means of ion thrusters (field emission electical 
propulsion, FEEP).

\item For drag-free control the satellite needs an appropriate 
reference sensor.

\item FEEPs are not effective for orbit heights less than 1000 km, 
because of the high gas density in lower orbits.

\item To avoid charging of the capacitive reference sensor by 
interactions with high energy protons, highly eccentric orbits, where the 
satellite passes the van--Allen--belt, are not appropriate.

\item The KT experiment requires a low orbit, because 
the experimental resolution depends directly on the satellite orbit 
velocity.

\item The precise attitude control requires a star sensor with a 
resolution of 10 arcsec.

\item Mechanical components for attitude control (e.g. fly wheel 
or mechanical gyros) cannot be used, because of the 
sensitivity of the experiment to vibrations.

\item Temperature stability requirements during integration times 
of more than 100 s can only be realized on orbits without or with rare 
eclipse intervals.

\item The thermal control of the satellie structure must achieve a stabilioty of 
$10^{-3}\hbox{K}/\sqrt{\hbox{Hz}}$.

\item The mission time is 6 months minimum.

\end{itemize}

OPTIS is designed to be launched on a micro--satellite with limited technological performance. 
Considering all experimental requirements, technological feasibility, 
launch capability,  and design philosophy the most 
feasible solution is to launch the satellite in a high elliptical 
orbit, attainable via a geo--transfer orbit (GTO) by lifting the 
perigee. In this scenario, the satellite is first launched as an 
auxiliary payload (ASAP5) by an ARIANE 5 rocket into the GTO with an 
apogee of 35\,800 km and a perigee of only 280 km. An additional 
kick--motor on the satellite will lift the satellite in its final 
orbit with a perigee of about 10\,000 km, corresponding to $\Delta 
v=0.75 \;\hbox{km/s}$. A minimum height of 10\,000 km enables one to 
use ion thrusters (FEEPs) and avoids flying through the van--Allen--belt. Also, 
the orbit eccentricity of $\epsilon
\simeq 0.41$ is high enough to attain sufficient velocity differences
for the KT experiment. Although the 
relatively high orbit reduces the in--orbit velocity of the satellite 
by a factor of 2.8 compared to a low earth circular orbit, it is still 20 times higher than for an earth--based experiment.

\subsection{Satellite bus structure and subsystems}

The satellite has cylindrical shape, with a height and a diameter of 
about 1.5 m. It spins around its cylindrical axis which is 
always directed to the sun. The front facing the sun is covered by solar 
panels. Behind the front plate serving as thermal shield a cylindrical 
box for on--board electronics and thruster control is located. 
The experimental box whose temperature is actively controlled is placed below 
the electronics. A stringer structure below the experimental box 
carries the kick--motor and the fuel tank. 4 clusters of ion thrusters 
for fine (drag--free) attitude control as well as 3 clusters of cold 
gas thrusters for coarse attitude control and first acquisition 
operation are mounted circumferentially. The structure elements of the 
entire bus have to satisfy the extreme requirement for passive thermal 
control. Their thermal expansion coefficient has to be less than 
$10^{-6}/\hbox{K}$.

The total satellite mass is about 250 kg including 90 kg of experimental payload. The total power budget is estimated to be less than 250 W. 

\subsubsection{Attitude and orbit control}

During experimental and safe mode coarse attitude and orbit control 
are based on a sun sensor (1 arcsec resolution) and a star sensor (10 
arcsec resolution). Fiber gyros are used for spin-- and 
de--spin--maneuvers and serve to control the cold gas thrusters. The 
fine attitude control, also called the drag--free control, must be 
carried out with an accuracy of $10^{-10}\;\hbox{m/s}^{2}$ within the 
signal bandwidth of $10^{-2}$ to $10^{-3}$ Hz, depending on the 
satellite's spin rate. The general principle of drag--free--control is 
to make the satellite's trajectory as close as 
possible a geodesic. Therefore, a capacitive reference sensor \cite{Touboul01} 
is used. The sensor unit consists of a test mass whose 
movements are measured capacitively with respect to all 6 degrees of 
freedom. Apart from the sensing electrodes, electrodes for servo--control 
surround the test mass and compensate its movement relative to 
the satellite structure. The signal is also used to control the satellite's 
movement via the ion thrusters. Thus, the test mass falls quasi--freely  
and is shielded against all disturbances by the satellite, in 
particular against solar pressure and drag. Because the servo-control 
is influenced by back--action effects, the test mass and the servo--control form 
a spring--mass system whose spring constant and eigenfrequnecy must be 
adapted to the signal bandwidth. 
Therefore, electrode surfaces, charging by external sources as 
well as the precision of the test mass and the electrode alignments influence 
the measurement of $x$ directly and make necessary repeated in--orbit 
calibration maneuvers \cite{Touboul01}. The chosen orbit avoids charging effects as much as possible.

FEEP ion thrusters (Field Emission Electric Propulsion) must be used 
to overcome (1) solar radiation pressure acting on the satellite and 
disturbing its free fall behaviour, and (2) to control the residual 
acceleration down to $10^{-10}\; \hbox{m/s}^2$ in the signal bandwidth. 
The first requirement sets an upper limit for the thrust: linear 
forces acting on the satellite are less than $50\; \mu \hbox{N}$ in all 
3 axes and maximum torques are about $10\; \mu \hbox{Nm}$. The second 
condition determines the resolution of thrust control which have to be 
done with an accuracy of about $0.1\; \mu \hbox{N}$. A FEEP able to 
satisfy these requirements is the Indium Liquid Metal Ion Source 
(LMIS) of the Austrian Research Centers Seibersdorf (ARCS) 
\cite{ARCS00}. Thrust is produced by accelerating indium ions in a 
strong electrical field. A sharpened tungsten needle is mounted in the 
centre of a cylindrical indium reservoir bonded to a ceramic tube 
which houses a heater element for melting the indium. Ion emission is 
started by applying a high positive potential between the tip covered 
with a thin indium film and an accelerator electrode. To avoid 
charging a neutralizer emitting electones is installed. The FEEPs have 
a length of 2 cm, a diameter of 4 mm, and weigh only several grams 
sufficient for the entire mission. A continuous thrust of $1.5\; \mu 
\hbox{N}$ per thruster is available. For a satellite diameter of about 
1.5 m (the maximum ASAP 5 size), the solar radiation pressure of ca. 
$4.4 \;\mu \hbox{N}/\hbox{m}^2$ and the radiation pressure of the earth 
albedo of $1.2\; \mu \hbox{N}/\hbox{m}^2$ sum up to a total drag of about $10 \;\mu \hbox{N}$. Considering thruster noise, misalignments and other disturbing effects, a 
continuous thrust of $12\; \mu \hbox{N}$ would be sufficient. To control 
all 6 degrees of freedom a minimum of 3 clusters of 4 thrusters is 
necessary. To guarantee a continuous thrust with some redundancy 4 
clusters are desirable. The power consumption is less than 3 W on 
average with peaks up to 25 W.

\section{Conclusion}

The proposed OPTIS mission is capable to make considerable improvements, up to three orders of magnitude, in the tests of SR and GR. 
It is designed to be a low cost mission. 
It is based on using (i) recent laboratory developments in optical technology and (ii) the advantages of space conditions: quit environment, long integration time, large velocities and large potential differences. 

The optical technology includes an optical comb generator, stabilized lasers and highly stable cavities. 
We remark that an alternative optical cavity system consisting of a monolithic silicon block operated at $\sim 140\;\hbox{K}$, the temperature where the thermal expansion coefficient vanishes, should be studied \cite{RichardHamilton91}. 
The advantage of this approach could be a significantly reduced level of creep and therefore a corresponding improvement of the KT test.

\end{document}